\documentstyle[12pt]{article}
\input epsf.sty

\catcode`@=11
\def\chkspace{%
  \relax   
  \begingroup\ifhmode\aftergroup\dochksp@ce\fi\endgroup}
\def\dochksp@ce{%
  \unskip              
  \futurelet\chkspct@k\d@chkspc  
}
\def\d@chkspc{%
  \let\nxtsp@ce=\relax
  \ifx\chkspct@k.\else     
    \ifx\chkspct@k,\else
      \ifx\chkspct@k;\else
        \ifx\chkspct@k!\else
          \ifx\chkspct@k?\else
            \ifx\chkspct@k:\else
              \ifx\chkspct@k)\else
              \ifx\chkspct@k(\else
                \ifx\chkspct@k]\else
                  \ifx\chkspct@k-\else
                    \ifx\chkspct@k\egroup\else  
                      \let\nxtsp@ce=\put@space  
                    \fi
                  \fi
                \fi
              \fi
              \fi
            \fi
          \fi
        \fi
      \fi
    \fi
  \fi
  \nxtsp@ce
}
\def\put@space{$\;$}
\catcode`@=12

\def\ra{{$\rightarrow$}\chkspace}
\def\etal{{\it et al.}\chkspace}

\def\adhoc{{\it ad hoc}\chkspace}

\def\eg{{\it eg.}\chkspace}

\def\apriori{{\it a priori}\chkspace}

\def\ep{{e$^+$e$^-$}\chkspace}

\def\gluino{\relax\ifmmode \tilde{g} \else $\tilde{g}$ \fi\chkspace}

\def\bb{\relax\ifmmode {\rm b}\bar{\rm b}
       \else ${\rm b}\bar{\rm b}$ \fi\chkspace}
\def\cc{\relax\ifmmode {\rm c}\bar{\rm c}
       \else ${\rm c}\bar{\rm c}$ \fi\chkspace}
\def\tt{\relax\ifmmode {\rm t}\bar{\rm t}
       \else ${\rm t}\bar{\rm t}$ \fi\chkspace}
\def\ss{\relax\ifmmode {\rm s}\bar{\rm s}
       \else ${\rm s}\bar{\rm s}$ \fi\chkspace}

\def\qqg{\relax\ifmmode {\rm q}\overline{\rm q}{\rm g}
\else q$\overline{\rm q}$g \fi\chkspace}

\def\afb{\relax\ifmmode A_{FB} \else
{{$A_{FB}$}}\fi\chkspace}
\def\afbb{\relax\ifmmode A_{FB}^b \else
{{$A_{FB}^b$}}\fi\chkspace}
\def\pafb{\relax\ifmmode \tilde{A}_{FB} \else
{{$\tilde{A}_{FB}$}}\fi\chkspace}
\def\pafbb{\relax\ifmmode \tilde{A}_{FB}^b \else
{{$\tilde{A}_{FB}^b$}}\fi\chkspace}

\def\pafbzo{\relax\ifmmode \tilde{A}_{FB}|_{O(0)} \else
{{$\tilde{A}_{FB}|_{O(0)}$}}\fi\chkspace}
\def\pafbfo{\relax\ifmmode \tilde{A}_{FB}|_{\oalp} \else
{{$\tilde{A}_{FB}|_{\oalp}$}}\fi\chkspace}
\def\pafbso{\relax\ifmmode \tilde{A}_{FB}|_{\oalpsq} \else
{{$\tilde{A}_{FB}|_{\oalpsq}$}}\fi\chkspace}
\def\pafbto{\relax\ifmmode \tilde{A}_{FB}|_{\oalpc} \else
{{$\tilde{A}_{FB}|_{\oalpc}$}}\fi\chkspace}

\def\pafbbzo{\relax\ifmmode \tilde{A}_{FB}^b|_{O(0)} \else
{{$\tilde{A}_{FB}^b|_{O(0)}$}}\fi\chkspace}
\def\pafbbfo{\relax\ifmmode \tilde{A}_{FB}^b|_{\oalp} \else
{{$\tilde{A}_{FB}^b|_{\oalp}$}}\fi\chkspace}
\def\pafbbso{\relax\ifmmode \tilde{A}_{FB}^b|_{\oalpsq} \else
{{$\tilde{A}_{FB}^b|_{\oalpsq}$}}\fi\chkspace}
\def\pafbbto{\relax\ifmmode \tilde{A}_{FB}^b|_{\oalpc} \else
{{$\tilde{A}_{FB}^b|_{\oalpc}$}}\fi\chkspace}

\def\afbo0{\tilde{A}_{FB}|_{O(0)}}
\def\afbo1{\tilde{A}_{FB}|_{\oalp}}
\def\afbo2{\tilde{A}_{FB}|_{\oalpsq}}
\def\afbo3{\tilde{A}_{FB}|_{\oalpc}}

\def\lam{\relax\ifmmode \Lambda_{\overline{MS}}
       \else {{$\Lambda_{\overline{MS}}$}}\fi\chkspace}
\def\lamuds{\relax\ifmmode \Lambda^{(3)}_{\overline{MS}}
       \else {{$\Lambda^{(3)}_{\overline{MS}}$}}\fi\chkspace}
\def\lamudsc{\relax\ifmmode \Lambda^{(4)}_{\overline{MS}}
       \else $\Lambda^{(4)}_{\overline{MS}}$\fi\chkspace}
\def\lamudscb{\relax\ifmmode \Lambda^{(5)}_{\overline{MS}}
       \else $\Lambda^{(5)}_{\overline{MS}}$\fi\chkspace}

\def\alp{\relax\ifmmode \alpha_s\else $\alpha_s$\fi\chkspace}
\def\alpbar{\relax\ifmmode \overline{\alpha_s}
       \else $\overline{\alpha_s}$\fi\chkspace}
\def\alpmz{\relax\ifmmode \alpha_s(M_Z)\else $\alpha_s(M_Z)$\fi\chkspace}
\def\alpmzsq{\relax\ifmmode \alpha_s(M_Z^2)
       \else $\alpha_s(M_Z^2)$\fi\chkspace}

\def\oalp{\relax\ifmmode O(\alpha_s)\else{{O($\alpha_s$)}}\fi\chkspace}
\def\oalpsq{\relax\ifmmode O(\alpha_s^2)
           \else{{O($\alpha_s^2$)}}\fi\chkspace}
\def\oalpc{\relax\ifmmode O(\alpha_s^3)
           \else{{O($\alpha_s^3$)}}\fi\chkspace}
\def\oalpf{\relax\ifmmode O(\alpha_s^4)
           \else{{O($\alpha_s^4$)}}\fi\chkspace}

\def\plb{Phys. Lett.\chkspace}
\def\npb{Nucl. Phys.\chkspace}

\def\prl{Phys. Rev. Lett.\chkspace}
\def\prd{Phys. Rev.\chkspace}
\def\zpc{Z. Phys.\chkspace}

\def\z0{{$Z^0$}\chkspace}
\def\Dst{\relax\ifmmode {\rm D}^* \else {D$^*$}\fi\chkspace}
\def\Dpl{\relax\ifmmode {\rm D}^+ \else {D$^+$}\fi\chkspace}
\def\D0{\relax\ifmmode {\rm D}^0 \else {D$^0$}\fi\chkspace}
\def\Kst{\relax\ifmmode {\rm K}^* \else {K$^*$}\fi\chkspace}
\def\K0{\relax\ifmmode {\rm K}^0_s \else {K$^0_s$}\fi\chkspace}
\def\Kpl{\relax\ifmmode {\rm K}^+ \else {K$^+$}\fi\chkspace}
\def\Kstz{\relax\ifmmode {\rm K}^{*0} \else {K$^{*0}$}\fi\chkspace}

\renewcommand{\baselinestretch}{1.5}

 1

\catcode`\@=11 
\makeatletter
\def\@seccntformat#1{\csname the#1\endcsname.\hskip 1em}

\makeatother
 
\begin{document}
 
\thispagestyle{empty}
\begin{flushright}
SLAC-PUB-7489\\
  June 1997\\
\end{flushright}
 
\vskip .5truecm
 
\begin{center}
 {\Large \bf MEASUREMENT OF THE}
 {\Large \bf B HADRON ENERGY DISTRIBUTION}
 {\Large \bf IN $Z^0$ DECAYS$^*$}
\normalsize
\vskip .5truecm
 
{\bf The SLD Collaboration$^{**}$}\\
Stanford Linear Accelerator Center \\
Stanford University, Stanford, CA~94309
 
\end{center}
 
\normalsize

\vskip .6truecm

\begin{center}
{\bf ABSTRACT }
\end{center}

\noindent 
We have measured the B hadron energy distribution in \z0 decays
using a sample of semi-leptonic B decays recorded
in the SLD experiment at SLAC. The energy of each tagged
B hadron was reconstructed using information
from the lepton and a partially-reconstructed charm-decay vertex.
We compared
the scaled energy distribution with several models
of heavy quark fragmentation. The average scaled energy of primary 
B hadrons was found to be
$<x_{E_B}> = 0.716\; \pm \; 0.011 ({\rm stat.})
\;^{+0.021}_{-0.022}\;({\rm syst.})$.
\vskip .4truecm
\centerline{\it Submitted to Physical Review D}
\vskip 1truecm

\vbox{\footnotesize\renewcommand{\baselinestretch}{1}\noindent
$^*$Work supported by Department of Energy
  contracts:
  DE-FG02-91ER40676 (BU),
  DE-FG03-91ER40618 (UCSB),
  DE-FG03-92ER40689 (UCSC),
  DE-FG03-93ER40788 (CSU),
  DE-FG02-91ER40672 (Colorado),
  DE-FG02-91ER40677 (Illinois),
  DE-AC03-76SF00098 (LBL),
  DE-FG02-92ER40715 (Massachusetts),
  DE-FC02-94ER40818 (MIT),
  DE-FG03-96ER40969 (Oregon),
  DE-AC03-76SF00515 (SLAC),
  DE-FG05-91ER40627 (Tennessee),
  DE-FG02-95ER40896 (Wisconsin),
  DE-FG02-92ER40704 (Yale);
  National Science Foundation grants:
  PHY-91-13428 (UCSC),
  PHY-89-21320 (Columbia),
  PHY-92-04239 (Cincinnati),
  PHY-95-10439 (Rutgers),
  PHY-88-19316 (Vanderbilt),
  PHY-92-03212 (Washington);
  The UK Particle Physics and Astronomy Research Council
  (Brunel and RAL);
  The Istituto Nazionale di Fisica Nucleare of Italy
  (Bologna, Ferrara, Frascati, Pisa, Padova, Perugia);
  The Japan-US Cooperative Research Project on High Energy Physics
  (Nagoya, Tohoku);
  The Korea Science and Engineering Foundation (Soongsil).}

\vfill
\eject
 
\section{Introduction}
 
The production of heavy hadrons (H) in \ep annihilation provides a
laboratory for the study of heavy-quark (Q) jet fragmentation. This is 
commonly characterised in terms of the observable 
$x_{E_H}$ $\equiv$ $2E_H/\sqrt{s}$, where
$E_H$ is the energy of a B or D hadron containing a b or c quark,
respectively, and $\sqrt{s}$ is the c.m. energy. In contrast to light-quark
jet fragmentation one expects~\cite{Bj} the distribution of $x_{E_H}$, 
$D(x_{E_H})$, to peak at an $x_{E_H}$-value significantly above 0. 
Since the hadronisation process is intrinsically non-perturbative $D(x_{E_H})$ 
cannot be calculated directly using perturbative Quantum Chromodynamics
(QCD). However, the distribution of the closely-related variable
$x_{E_Q}$ $\equiv$ 2$E_Q/\sqrt{s}$ can be calculated
perturbatively \cite{MN,DKT,BCFY} and related, via model-dependent
assumptions, to the observable quantity $D(x_{E_H})$; a number of such
models of heavy quark fragmentation have been proposed
\cite{bowler,pete,lund}. Measurements of $D(x_{E_H})$ thus serve to
constrain both perturbative QCD and the model predictions. 
Furthermore, the measurement of $D(x_{E_H})$ at different c.m. energies
can be used to test QCD evolution, and comparison of $D(x_{E_B})$
with $D(x_{E_D})$ can be used to test heavy quark symmetry~\cite{Lisa}. 
Finally, the uncertainty on the forms of $D(x_{E_D})$ and $D(x_{E_B})$
must be taken into account in studies of the production and decay of heavy
quarks, see \eg~\cite{heavy}; more accurate measurements of these forms 
will allow increased precision in tests of the electroweak heavy-quark sector.

Here we consider measurement of the B hadron scaled energy distribution
$D(x_{E_B})$ in $Z^0$ decays. Earlier studies \cite{early} 
used the momentum spectrum of the lepton from semi-leptonic B decays to 
constrain the mean value $<x_{E_B}>$ and found it to be approximately
$0.70$; this is in agreement with the results of similar studies at $\sqrt{s}$
= 29 and 35 GeV~\cite{petra}. In more recent analyses~\cite{alshape,shape} 
the scaled energy distribution 
$D(x_{E_B})$ has been measured by reconstructing B hadrons via their
B \ra D$l$X decay mode; we have applied a similar technique.  
We used the precise SLD tracking system to select jets containing a
B \ra D$l$X decay, where
the charmed hadron D was identified semi-inclusively from a secondary decay
vertex formed from charged tracks. Each hadronic vertex was then associated 
with a lepton $l$ 
($l=e$ or $\mu$) with large momentum transverse to the jet direction.
Neutral energy depositions measured in the hermetic calorimeter, 
as well as the energies of charged tracks, that were not
associated with the D$l$ system were subtracted from the jet energy
to yield the reconstructed B hadron energy. 
This measurement technique may be useful to 
B-lifetime or B-mixing analyses~\cite{mix} where
the proper time $t=L/\sqrt{\gamma^2-1}$, where $\gamma=E_B/m_B$, 
$m_B$ is the B hadron mass and $L$ is the decay length, 
must be known accurately.
We then compared the B energy distribution with the perturbative QCD and
phenomenological model predictions. 
 
\section{Apparatus and Hadronic Event Selection}
 
The e$^+$e$^-$ annihilation events produced at the $Z^0$ resonance
by the SLAC Linear Collider (SLC)
were recorded using the SLC Large Detector (SLD).
A general description of the SLD can be found elsewhere~\cite{sld}.
This analysis used charged tracks measured in the Central Drift
Chamber (CDC)~\cite{cdc} and in the Vertex Detector (VXD)~\cite{vxd},
energy clusters measured in the
Liquid Argon Calorimeter (LAC)~\cite{lac}, and muons measured in the
Warm Iron Calorimeter (WIC)~\cite{wic}.
Electron identification utilizes CDC tracks and LAC clusters~\cite{sldlept}.

Momentum measurement is provided by a uniform axial magnetic field of 0.6~T.
The CDC and VXD  give a momentum resolution of
$\sigma_{p_{\perp}}/p_{\perp}$ = $0.01 \oplus 0.0026p_{\perp}$,
where $p_{\perp}$ is the track momentum transverse to the beam axis in
GeV/$c$. Including
the uncertainty on the primary interaction point (IP), the resolution on the 
charged-track impact parameter ($d$) projected in the plane perpendicular
to the beamline is $\sigma_d$ =
11$\oplus$70/$(p_{\perp} \sqrt{\sin\theta})$ $\mu$m, where
$\theta$ is the polar angle with respect to the
beamline. This results in a mean resolution on reconstructed 2-prong 
vertices (Section 3) of $\sigma_{V_{\parallel(\perp)}} = 400\: (25)$ $\mu$m
for the projection on an axis along
(perpendicular to) the vertex flight direction.
The LAC electromagnetic energy scale was calibrated from
the measured $\pi^0$ \ra $\gamma\gamma$ signal~\cite{saul,church};
the electromagnetic energy resolution is $\sigma_E/E\approx
0.15/\sqrt{E({\rm GeV})}$. 
 
The trigger and initial selection of hadronic events are described
in~\cite{alr}.
A set of cuts was applied to the data to select well-measured tracks
and events well-contained within the detector acceptance.
Charged tracks were required to have a distance of
closest approach transverse to the beam axis within 5 cm,
and within 10 cm along the axis from the measured interaction point,
as well as $|\cos \theta |< 0.80$, and $p_\perp > 0.15$ GeV/c.
Events were required to have a minimum of seven such tracks,
a thrust axis~\cite{thrust} polar angle $\theta_T$
within $|\cos\theta_T|<0.70$, and
a charged visible energy $E_{vis}$ of at least 20~GeV,
which was calculated from the selected tracks assigned the charged pion
mass. From our 1993-95 data sample 108650 events passed these cuts.
The efficiency  for selecting hadronic events satisfying
the $|\cos \theta_T |$ cut was estimated to be above $96\%$.
The background in the selected event sample was estimated to be
$0.1\pm 0.1\%$, dominated by $Z^0 \rightarrow \tau^+ \tau^-$ events.

Calorimeter clusters used in the subsequent jet-finding analysis (Section 4)
were required to comprise at least two calorimeter towers, each containing
an energy of at least 100 MeV, and to have a total energy greater than 250 MeV.
Electromagnetic clusters used in the non-B-associated neutral energy
measurement were further required to have less than the smaller of,
25\% of their energy and 600 MeV, in the hadronic section of the LAC. 

The efficiency for reconstructing B hadrons, the background in the selected
sample, and the resolution of the method were evaluated (Sections 3 and 4) 
using a detailed Monte Carlo (MC) simulation. The
JETSET 7.4~\cite{jetset74} event generator was used, with parameter
values tuned to hadronic \ep annihilation data~\cite{tune},
combined with a simulation of B-decays
tuned to $\Upsilon(4S)$ data~\cite{sldsim} and a simulation of the SLD
based on GEANT 3.21~\cite{geant}.
Inclusive distributions of single particle and event topology observables
in hadronic events were found to be well-described by the
simulation~\cite{PRD.Ohnishi}. 
There is now evidence that roughly 21\% of all promptly-produced B hadrons in
\z0 \ra \bb events are B$^{**}$ mesons~\cite{Bstst}; since JETSET does not
produce B$^{**}$ mesons we have corrected the simulation to account for them.
Using an event weighting technique we produced a generator-level distribution 
of B hadron energies in which the energy $E_B$ of 20.7\% of all  
B hadrons was adjusted to be $E_B-E_{\pi}$, where the pion
energy $E_\pi$ was produced according to an isotropic 2-body decay
distribution for B$^{**}\rightarrow$ B$\pi^\pm$, assuming a
B$^{**}$ mass of 5.7 GeV/$c^2$. Uncertainties in this simulation of B$^{**}$
production were taken into account in the systematic errors (Section 7). 
 
\section{B Hadron Selection}
 
Hadronic events were required to contain a lepton candidate
within the barrel tracking system with $|{\rm cos}\theta|<0.7$. 
We then applied the JADE jet-finding algorithm~\cite{JADE} to 
the LAC clusters in each selected event to define a jet topology.
With a jet-resolution criterion of
$y_{c}=0.07$, 82.9\% of the events were classified as 2-jet-like and
17.1\% as 3-jet-like. 
Kinematic information based on this
topological classification was used subsequently (Section 4) in the
calculation of the B hadron energy. Events in which the lepton had a
transverse momentum w.r.t. its jet axis, $p_t$, of at least 1 GeV/$c$
were retained for further analysis. In jets containing more than one such
lepton only the highest-$p_t$ lepton was labelled for association with a
D vertex and any lower-momentum leptons were used in the D-vertex-finding.
 
In each selected jet we then searched for a 
secondary D vertex among the non-lepton tracks. Tracks were required to 
comprise at least 40 CDC hits and one VXD hit, to be well contained 
within the CDC with $|\cos\theta|\le 0.70$, 
to have momentum in the range $0.15<p<55$ GeV/$c$, and to have a
transverse impact parameter, normalised by its error, of
$d/\sigma_d$ $>$ 1. Tracks
from $K_s^0$ and $\Lambda^0$ decays and $\gamma$ conversions were suppressed by
requiring the distance of closest approach to the IP
in the planes both perpendicular to, and containing, the beamline to be 
less than 1 cm. 
Two-prong vertices were first formed from all pairs of
tracks whose distance-of-closest-approach was less than $0.012$ cm and whose
fit to a vertex satisfied $\chi^2$ $<5$. 
A multi-prong D-vertex candidate was then defined to comprise the tracks in all
accepted two-prong vertices in the jet, and 
to be located at the position of the two-prong vertex containing the
track with the largest normalised transverse impact parameter $d/\sigma_d$.

The tracks in each D vertex were each assigned the charged pion mass and were 
then combined by adding their four-vectors to obtain the
vertex invariant mass, $m_D$, and the vertex momentum vector. 
The vertex flight distance from the IP
was projected onto the jet axis to obtain the quantity $r_D$. 
Events were retained if at least one jet 
contained a D vertex with $0.3<m_D<1.9$ GeV/$c^2$, 
$r_D$ $>$ 0.05 cm, $r_D$ normalised by its error larger than unity, and  
the distance-of-closest-approach between the  
lepton track and the extrapolated D-vertex momentum vector 
was less than $0.012$ cm. The lepton and D-vertex tracks were then fitted to
a common candidate B vertex. The combined D-vertex and lepton
invariant mass, $m_B$, and the projection of the vector between the
B- and D-vertex positions onto the D-vertex momentum vector, $r_B$, were 
calculated. Events were selected in which
$m_B<4.5$ GeV/$c^2$, $r_B$ $>$ $0.025$ cm, and $r_B$ normalized by its 
error was larger than unity.

For the selected events, 
distributions of the number of tracks per D vertex, $N_D$, and of $m_D$, $r_D$,
$m_B$, and  $r_B$ are shown in Fig.~1. Also shown are the simulated
distributions in which the contribution from selected true B \ra D$l$X decays 
is indicated. In Fig.~2 the distributions                               
of lepton transverse momentum with respect to the jet axis, $p_t$, are shown
for candidates passing all cuts except the requirement that $p_t$ 
be above 1 GeV/$c$; the simulated distributions are also shown, and the 
contributions from different processes are indicated. The final sample 
comprises 597 events, 293 in the muon, and 304 in the electron, channels. 
Using the simulation we estimate that 
the purity of this sample, defined to be the fraction of the tagged
events whose identified leptons $l$ are from true B \ra D$l$X decays, 
is 69.2\%; a further 18\% of the selected events contain B decays with a
cascade, punch-through or mis-identified lepton, and are still useful. 
The estimated composition of the \bb events in terms of the B hadron species is 
shown in 
Table~1. The remaining 12.8\% of the event sample comprises non-\bb events. 
The efficiency for selecting B hadron decays in the selected hadronic event
sample is shown, as a function of
$x_{E_B}$, in Fig.~3; the overall efficiency is 1.1\%.
 
\section{Measurement of the B Energies}
 
In each selected event we first defined the jet energies
by using kinematic information. The
2-jet events were divided into two hemispheres by the plane normal to
the thrust axis and the jet in each hemisphere was assigned the beam
energy. For the 3-jet events we corrected
the jet energies according to the angles between the jet axes, assuming
energy and momentum conservation and massless kinematics. Labelling the
jets arbitrarily 1, 2 and 3, and the corresponding inter-jet angles
$\theta_{23}$, $\theta_{13}$ and $\theta_{12}$,
the corrected energy of jet 1 is given by:
\begin{eqnarray}
E_{1}=\sqrt{s}(\sin\theta_{23})/
                 (\sin\theta_{12}+\sin\theta_{23}+\sin\theta_{13}),
\end{eqnarray}
with corresponding expressions for jets 2 and 3. This procedure results in
improved jet energy resolution. 
 
We then proceeded to reconstruct the B hadron energy $E_B^{Rec}$:
\begin{eqnarray}
E_B^{Rec} \quad = \quad E_{jet} - E_{frag}, \label{eq:energy}
\end{eqnarray}
where $E_{jet}$ is the energy of the jet containing the candidate B vertex and
$E_{frag}$ is the energy in the same jet that is not attributed to the B,
\begin{eqnarray}
E_{frag}\quad = \quad f^{chg}\; E_{frag}^{chg} + f^{neu}\;E_{frag}^{neu} 
\label{eq:efrag}
\end{eqnarray}
where $E^{chg}_{frag}$ and $E^{neu}_{frag}$ are the measured charged and neutral
energy components respectively, and $f^{chg}$ and $f^{neu}$ are correction
factors described below. We define
$E^{chg}_{frag}$ to be the sum of the energy, using the momentum and
assuming the pion mass, of all the charged
tracks in the jet excluding the candidate B-vertex
tracks; $E^{neu}_{frag}$ is defined to be the
sum of the energy of the electromagnetic calorimeter clusters in the jet
that are not associated with charged tracks. A cluster was defined as
unassociated if it had no charged track extrapolating to it to within an angle 
4$\sigma_{cl}$ from its centroid, where
$\sigma_{cl}$ = $\sqrt{{\sigma_{cl}^{\theta}}^2+{\sigma_{cl}^{\phi}}^2}$
and $\sigma_{cl}^{\theta}$ and $\sigma_{cl}^{\phi}$ are the measured cluster 
widths in polar- and azimuthal-angle, respectively. 
The distributions of $E^{chg}_{frag}$ and $E^{neu}_{frag}$ are shown
in Fig.~4.
 
This procedure will \apriori misassign the energy of any unassociated 
neutral particle from the D decay to the non-B energy $E_{frag}$.
Similarly, the energy of any charged
track from the D decay that is not associated with the reconstructed D vertex
will be misassigned to $E_{frag}$.
We have used our MC simulation to study these effects and show in
Fig.~5 the correlation
between the reconstructed and true values of $E^{neu}_{frag}$ and
$E^{chg}_{frag}$. As expected, both the charged and neutral components are
typically slightly overestimated by the reconstruction method.
We fitted an \adhoc second-order polynomial to each
correlation to determine an average energy-dependent correction factor,
$f^{chg}$ ($f^{neu}$) (Eq.~3), which we applied to the non-B
charged (neutral) energy component $E_{frag}^{chg}$ ($E_{frag}^{neu}$)  
of each tagged
jet in the data sample. Uncertainties in these corrections were 
included in the systematic errors (Section~7).

We have used our simulation to
estimate the resolution of the method for reconstructing the B hadron energy. 
We compared the reconstructed scaled B energy $x_{E_B}^{rec}$
with the input scaled energy $x_{E_B}^{true}$ and show in
Fig.~6 the distribution of the quantity 
$(x_{E_B}^{true}-x_{E_B}^{rec})/x_{E_B}^{true}$. The resolution may be
characterised by a parametrisation comprising the sum of two Gaussian
distributions. The result of such a fit, in which the Gaussian 
centers, normalisations and widths were allowed to vary, is shown in Fig.~6.
The narrower Gaussian of width $\sigma$ = 0.10
represents 65\% of the fitted area, and the wider Gaussian of width
$\sigma$ = 0.33 represents the remainder. It can be seen from Fig.~6 that the
population corresponding to the `inner core' is somewhat underestimated by this
technique since the parametrisation does not describe the central bin.
We repeated this exercise in subset regions of $x_{E_B}^{true}$ and found
the inner core resolution (population) to be 
0.27 (84\%) for $0.0<x_{E_B}^{true}<0.6$, 
0.09 (70\%) for $0.6<x_{E_B}^{true}<0.8$, and
0.06 (79\%) for $0.9<x_{E_B}^{true}<1.0$; as expected the resolution 
is better for more energetic B hadrons. 
Choosing the bin width to be roughly half of our mean resolution we show
the measured distribution of $x_{E_B}^{rec}$, $D^{data}(x_{E_B}^{rec})$, 
in Fig.~7.
Also shown in this figure is the simulated distribution in which the background 
contribution from non-\bb events is indicated. 
  
\section{Comparison with Model Predictions}

It is interesting to compare our measured B hadron energy distribution
with the theoretical predictions. The event generator used in our 
simulation is 
based on a perturbative QCD `parton shower' for production of quarks and
gluons, together with the phenomenological Peterson function~\cite{pete} 
(Table 2) to account for the fragmentation of b and c quarks into B and 
D hadrons, respectively,  
within the iterative Lund string hadronisation mechanism~\cite{jetset74}; 
this simulation yields a generator-level 
primary B-hadron energy distribution with $<x_{E_B}>$ = 0.693\footnote{
We used a value of the Peterson function 
parameter $\epsilon_b$ = 0.006~\cite{sldrb}.}.
It is apparent (Fig.~7) that this
simulation does not reproduce the data well; the $\chi^2$ 
for the comparison is 36.7 for 15 bins.

We have also considered alternative forms of the fragmentation function
based on the phenomenological model of the Lund group~\cite{lund}, the
perturbative QCD calculations of Braaten \etal~\cite{BCFY}, (BCFY) and of
Nason \etal~\cite{MN} (NCM), as well as \adhoc parametrisations 
based on a function used by the ALEPH Collaboration~\cite{alshape} and on
a third-order polynomial. These functions are listed in Table~2. 

In order to make a consistent comparison of each function
with the data we adopted the following procedure. Starting values of
the arbitrary parameters were assigned and the corresponding distribution 
of scaled primary B hadron energies, $D^{MC}(x_{E_B}^{true})$, 
was reproduced in our MC-generated \bb event sample, 
{\it before} simulation of the detector, by weighting events accordingly. 
The resulting distribution, after simulation of the detector,
application of the analysis cuts and background subtraction, 
of reconstructed B hadron energies,
$D^{MC}(x_{E_B}^{rec})$, was then compared with the background-subtracted 
data distribution and the $\chi^2$ 
value was
calculated. This process was iterated to find the minimum in $\chi^2$, yielding
a parameter set that gives an optimal description of the reconstructed
data by the input fragmentation
function. This procedure was applied for each function listed in Table~2. 
The fitted parameters and minimum $\chi^2$ values are listed
in Table~3, and the corresponding $D^{MC}(x_{E_B}^{rec})$
are compared with the data in Fig.~8. Each function reproduces the data.
We conclude that, within our resolution and with our current data
sample, we are unable to distinguish between these functions. It should be
noted, however, that the optimal third-order polynomial function has a small
negative minimum point in the region around $x_{E_B}^{true}=0.2$; since this
behaviour is unphysical we did not consider this function further in the
analysis.
 
\section{Correction of the B Energy Distribution}

In order to compare our results with those from other experiments it is
necessary to correct the reconstructed scaled B hadron energy distribution 
$D^{data}(x_{E_B}^{rec})$ for the
effects of non-B backgrounds, detector acceptance, event selection and
analysis bias, and initial-state radiation, as well as for bin-to-bin
migration effects caused by the finite resolution of the detector and the
analysis technique. We also corrected for the effects of B$^{**}$ decays
(Section 2) to derive the primary B hadron energy distribution.
We applied a $15\times15$ matrix unfolding procedure 
to $D^{data}(x_{E_B}^{rec})$ to obtain an estimate of the true distribution 
$D^{data}(x_{E_B}^{true})$:
\begin{eqnarray}
D^{data}(x_{E_B}^{true})\quad=\quad \epsilon^{-1}(x_{E_B}^{true}) \cdot 
E(x_{E_B}^{true},x_{E_B}^{rec}) \cdot (D^{data}(x_{E_B}^{rec})
-S(x_{E_B}^{rec})) 
\end{eqnarray}
where $S$ is a vector representing the background contribution, $E$ is a
matrix to correct for bin-to-bin migrations, and $\epsilon$ is
a vector representing the efficiency for selecting true B hadron
decays for the analysis. 

The matrices $S$, $E$ and $\epsilon$ were calculated from our MC
simulation; the elements of $\epsilon$ are shown in Fig.~3.
The matrix $E$ incorporates a
convolution of the input fragmentation function with the resolution of the
detector. We used 
in turn the Peterson, Lund, BCFY, NCM and ALEPH functions, with the
optimised parameters listed in Table~3, to produce both a generator-level 
input primary B hadron energy distribution $D^{MC}(x_{E_B}^{true})$, and a 
reconstructed distribution $D^{MC}(x_{E_B}^{rec})$, 
as discussed in the previous section. In each case
$E$ was evaluated by examining 
the population migrations of true B hadrons between bins of the input 
scaled B energy, $x_{E_B}^{true}$, and 
the reconstructed scaled B energy, $x_{E_B}^{rec}$. 

The data were then unfolded
according to Eq.~(4) to yield $D^{data}(x_{E_B}^{true})$, 
which is shown for each input fragmentation function in Fig.~9.
It can be seen that the shapes of $D^{data}(x_{E_B}^{true})$ differ 
systematically among the assumed input fragmentation functions. 
These difference were used to assign systematic errors, as
discussed in the next section.

\section{Systematic Errors}

We have considered sources of systematic uncertainty that potentially affect 
our measurement of
the B-hadron energy distribution. These may be divided into uncertainties in
modelling the detector and uncertainties on experimental measurements serving as
input parameters to the underlying physics modelling. 
For these studies our standard simulation, employing 
the Peterson fragmentation function, was used.

The uncertainty on the
correction of the non-B neutral jet energy component $E^{neu}_{frag}$
(Section 4) was estimated by changing 
the LAC cluster-energy selection requirement from 100 to 200 MeV, and by 
varying the LAC electromagnetic energy scale within our estimated 
uncertainty of $\pm$2.2\% of its nominal value~\cite{saul}. 
In each case
the difference in results relative to our standard procedure was
taken as the systematic uncertainty. 
A large source of detector modelling uncertainty was found to relate to
knowledge of the charged tracking efficiency of the
detector, which we varied by our estimated uncertainty of $\pm2.4$\%. 
In addition, in each bin of $x^{rec}_{E_B}$,
we varied the estimated contribution from fake leptons in the data sample
(Fig.~2) by $\pm$25\%. These uncertainties were 
assumed to be uncorrelated and were added in quadrature to obtain the 
detector modelling uncertainty in each bin of $x_{E_B}$.
                             
As a cross-check we also varied the event selection requirements.
The thrust-axis containment cut was varied in the range 
$0.65<|\cos\theta_T| < 0.70$, the minimum number of charged tracks
required was increased from 7 to 8, and the total charged-track energy
requirement was increased from 20 to 22 GeV.
In each case results consistent with the standard selection were obtained. 
As a further cross-check on jet axis modelling we systematically varied $y_c$ 
in the range $0.01\le y_c\le 0.15$ and repeated the analysis; results consistent
with the standard analysis were obtained.

A large number of measured quantities relating to the production and decay
of charm and bottom hadrons are used as input to our simulation. 
In \bb events we have considered the uncertainties on: 
the branching fraction for \z0 \ra \bb;
the rates of production of B$_u$, B$_d$ and B$_s$ mesons, and B baryons;
the rate of production of B$^{**}$ mesons, and the B$^{**}$ mass;
the branching ratios for B \ra D$^*$ and B \ra D$^{**}$;
the lifetimes of B mesons and baryons;
and the average charged multiplicity of B hadron decays.
In \cc events we have considered the uncertainties on: 
the branching fraction for \z0 \ra \cc;
the charmed hadron fragmentation function;
the rates of production of D$^0$, D$^+$ and D$_s$ mesons, and charmed baryons;
and the charged multiplicity of charmed hadron decays.
We have also considered the rate of production of \ss in the jet fragmentation
process, and the production of secondary \bb and \cc from gluon splitting.
The world-average values~\cite{heavy,sldrb} of these quantities used in our
simulation, as well as the respective uncertainties, are listed in
Table~4. 

The variation of each quantity within its uncertainty was produced in turn
in our simulated event sample using an event weighting technique \cite{sldrb}.
The matrices $S$ and $E$ (Section 6) were then reevaluated using
the simulated events, and the data were recorrected.
In each case the deviation w.r.t. the standard corrected result was taken as
a separate systematic error. These uncertainties were conservatively
assumed to be uncorrelated and were added in quadrature to obtain a total
physics modelling uncertainty in each bin of $x_{E_B}$.
                             
The model-dependence of the unfolding procedure was estimated by considering
the envelope of the unfolded results illustrated in Fig.~9. In each bin of
$x_{E_B}$ we calculated the average value of the five unfolded results, as 
well as the r.m.s. deviation. The average value was taken as our central
value in each bin, and the r.m.s. value was assigned as the respective
unfolding uncertainty.

\section{Summary and Conclusions}

We have used the precise SLD tracking system to reconstruct the energies of 
B hadrons in \ep \ra \z0 events via the B \ra D$l$X decay mode.
We estimate our resolution on the B energy to be about 10\% for roughly
65\% of the reconstructed decays. 
The distribution of reconstructed scaled B hadron energy, $D(x^{rec}_{E_B})$, 
was compared with perturbative QCD and phenomenological model predictions; the
calculations of Braaten, Cheung and Yuan and of Nason, Colangelo and Mele are
consistent with our data, as are the phenomenological models of Peterson
\etal and of the Lund group.
The distribution was then corrected for
bin-to-bin migrations caused by the resolution of the method and
for selection efficiency, as well as for the effects of B$^{**}$ production, to
derive the energy distribution of primary B hadrons produced by \z0 decays. 
Systematic uncertainties in the correction were considered.
The final corrected $x_{E_B}$ distribution $D(x_{E_B})$ is listed in Table~5
and shown in Fig.~10; 
the statistical, experimental systematic, and unfolding uncertainties are
indicated separately.

It is conventional to evaluate the mean of this distribution, $<x_{E_B}>$.
For each of the five functions used to correct the data we evaluated $<x_{E_B}>$
from the distribution that corresponds to the optimised parameters; 
these are listed
in Table~3. We took the average of the five values of $<x_{E_B}>$ as our
central result, and defined the unfolding uncertainty to be the r.m.s. 
deviation. We list in Table~4 the errors on $<x_{E_B}>$ resulting from
the study of detector and physics modelling described in Section~7. 
We obtained:
$$
<x_{E_B}> \quad = \quad 0.716\; \pm \; 0.011 ({\rm stat.})
\;^{+0.009}_{-0.011}\;({\rm exp.\;syst.})\; \pm\; 0.019 \;({\rm unfolding}),
$$
where the systematic error is the sum in quadrature of the 
individual contributions listed in Table~4.
It can be seen that $<x_{E_B}>$ 
is relatively insensitive to the variety of allowed forms of the shape of the
fragmentation function $D(x_{E_B})$.

Our results are in agreement with a previous measurement of the shape of 
the primary B hadron energy distribution at the \z0 resonance~\cite{alshape},
as well as with measurements of the shape~\cite{shape} and mean 
value~\cite{early} of the distribution for weakly-decaying B hadrons, 
after taking account of our estimate that the latter $<x_{E_B}>$ value is  
about 0.015 lower. Combining all systematic errors in quadrature we
obtain $<x_{E_B}>$ = $0.716\; \pm \; 0.011$ (stat.)
$\;^{+0.021}_{-0.022}$ (syst.).

\vfill
\eject

\vskip 2truecm

\section*{$^{**}$List of Authors}

\begin{center}
%
%
%
  \def\iADEL{$^{(1)}$}
  \def\iBOL{$^{(2)}$}
  \def\iBU{$^{(3)}$}
  \def\iBRUN{$^{(4)}$}
  \def\iUCSB{$^{(5)}$}
  \def\iUCSC{$^{(6)}$}
  \def\iCIN{$^{(7)}$}
  \def\iCSU{$^{(8)}$}
  \def\iCOLO{$^{(9)}$}
  \def\iCOL{$^{(10)}$}
  \def\iFER{$^{(11)}$}
  \def\iFRA{$^{(12)}$}
  \def\iILL{$^{(13)}$}
  \def\iLBL{$^{(14)}$}
  \def\iMIT{$^{(15)}$}
  \def\iMASS{$^{(16)}$}
  \def\iMISS{$^{(17)}$}
  \def\iMOSC{$^{(18)}$}
  \def\iNAG{$^{(19)}$}
  \def\iOREG{$^{(20)}$}
  \def\iPAD{$^{(21)}$}
  \def\iPERU{$^{(22)}$}
  \def\iPISA{$^{(23)}$}
  \def\iRUT{$^{(24)}$}
  \def\iRAL{$^{(25)}$}
  \def\iSOGANG{$^{(26)}$}
  \def\iSOONG{$^{(27)}$}
  \def\iSLAC{$^{(28)}$}
  \def\iTENN{$^{(29)}$}
  \def\iTOH{$^{(30)}$}
  \def\iVAND{$^{(31)}$}
  \def\iWASH{$^{(32)}$}
  \def\iWISC{$^{(33)}$}
  \def\iYALE{$^{(34)}$}
  \def\dead{$^{\dag}$}
  \def\andgen{$^{(a)}$}
  \def\andper{$^{(b)}$}
%
%
\mbox{K. Abe                 \unskip,\iNAG}
\mbox{K. Abe                 \unskip,\iTOH}
\mbox{T. Akagi               \unskip,\iSLAC}
\mbox{N.J. Allen             \unskip,\iBRUN}
\mbox{W.W. Ash               \unskip,\iSLAC$^\dagger$}
\mbox{D. Aston               \unskip,\iSLAC}
\mbox{K.G. Baird             \unskip,\iRUT}
\mbox{C. Baltay              \unskip,\iYALE}
\mbox{H.R. Band              \unskip,\iWISC}
\mbox{M.B. Barakat           \unskip,\iYALE
% ...
}
\it
  \vskip \baselineskip                   
  \vskip \baselineskip                   
%
%
%
  \iADEL
     Adelphi University,
     Garden City, New York 11530 \break
  \iBOL
     INFN Sezione di Bologna,
     I-40126 Bologna, Italy \break
  \iBU
     Boston University,
     Boston, Massachusetts 02215 \break
  \iBRUN
     Brunel University,
     Uxbridge, Middlesex UB8 3PH, United Kingdom \break
  \iUCSB
     University of California at Santa Barbara,
     Santa Barbara, California 93106 \break
  \iUCSC
     University of California at Santa Cruz,
     Santa Cruz, California 95064 \break
  \iCIN
     University of Cincinnati,
     Cincinnati, Ohio 45221 \break
  \iCSU
     Colorado State University,
     Fort Collins, Colorado 80523 \break
  \iCOLO
     University of Colorado,
     Boulder, Colorado 80309 \break
  \iCOL
     Columbia University,
     New York, New York 10027 \break
  \iFER
     INFN Sezione di Ferrara and Universit\`a di Ferrara,
     I-44100 Ferrara, Italy \break
  \iFRA
     INFN  Lab. Nazionali di Frascati,
     I-00044 Frascati, Italy \break
  \iILL
     University of Illinois,
     Urbana, Illinois 61801 \break
  \iLBL
     E.O. Lawrence Berkeley Laboratory, University of California,
     Berkeley, California 94720 \break
  \iMIT
     Massachusetts Institute of Technology,
     Cambridge, Massachusetts 02139 \break
  \iMASS
     University of Massachusetts,
     Amherst, Massachusetts 01003 \break
  \iMISS
     University of Mississippi,
     University, Mississippi  38677 \break
  \iMOSC
    Moscow State University,
    Institute of Nuclear Physics
    119899 Moscow, Russia    \break
  \iNAG
     Nagoya University,
     Chikusa-ku, Nagoya 464 Japan  \break
  \iOREG
     University of Oregon,
     Eugene, Oregon 97403 \break
  \iPAD
     INFN Sezione di Padova and Universit\`a di Padova,
     I-35100 Padova, Italy \break
  \iPERU
     INFN Sezione di Perugia and Universit\`a di Perugia,
     I-06100 Perugia, Italy \break
  \iPISA
     INFN Sezione di Pisa and Universit\`a di Pisa,
     I-56100 Pisa, Italy \break
  \iRUT
     Rutgers University,
     Piscataway, New Jersey 08855 \break
  \iRAL
     Rutherford Appleton Laboratory,
     Chilton, Didcot, Oxon OX11 0QX United Kingdom \break
  \iSOGANG
     Sogang University,
     Seoul, Korea \break
  \iSOONG
     Soongsil University,
     Seoul, Korea  156-743 \break
  \iSLAC
     Stanford Linear Accelerator Center, Stanford University,
     Stanford, California 94309 \break
  \iTENN
     University of Tennessee,
     Knoxville, Tennessee 37996 \break
  \iTOH
     Tohoku University,
     Sendai 980 Japan \break
  \iVAND
     Vanderbilt University,
     Nashville, Tennessee 37235 \break
  \iWASH
     University of Washington,
     Seattle, Washington 98195 \break
  \iWISC
     University of Wisconsin,
     Madison, Wisconsin 53706 \break
  \iYALE
     Yale University,
     New Haven, Connecticut 06511 \break
  \dead
     Deceased \break
  \andgen
     Also at the Universit\`a di Genova \break
  \andper
     Also at the Universit\`a di Perugia \break
\rm
%

\end{center}

\vfill\eject



\begin{table}[t]
\renewcommand{\baselinestretch}{1.0}  
\centering
\baselineskip=13pt
\medskip
\begin{tabular}{|c|c|c|} \hline
B species & $C$ (\%) & $\epsilon$ (\%) \\ \hline
B$_u$ &43 &92   \\
B$_d$ & 43 &87  \\
B$_s$ & 10 &89  \\
B baryons & 4 & 87  \\  \hline
\end{tabular}
\caption{
The composition $C$ of true B \ra D$l$X decays in the final sample; 
$\epsilon$ is the fraction of each species whose D vertices are correctly
reconstructed.
In all cases the MC statistical errors are less than 2\%.}
\end{table}


\begin{table}[t]
\renewcommand{\baselinestretch}{1.0}  
\medskip
\centering
\begin{tabular}{|l|l|c|} \hline
Function Name & Functional form $D(x)$ & Reference \\
\hline
Peterson & $\frac{1}{x}(1-\frac{1}{x}-\frac{\epsilon_b}{1-x})^{-2}$ &
 \cite{pete} \\
Lund & $\frac{1}{x}(1-x)^a \exp(-bm_T^2/x)$ & \cite{lund} \\
BCFY & 
$\frac{x(1-x^2)}{(1-(1-r)x)^6} [3 - xf_1(r) + x^2 f_2(r) - x^3 f_3(r) + x^4
f_4(r)]$
& \cite{BCFY}      \\
NCM & $\int dy\, g(x,y)y^\alpha(1-y)^\beta$& \cite{MN}\\
ALEPH & $\frac{1+b(1-x)}{x}(1-\frac{c}{x}-\frac{d}{1-x})^{-2}$&
 \cite{alshape}\\
3$^{rd}$-order Polynomial &  $1 + bx +cx^2 +dx^3 $ & \\  \hline
\end{tabular}
\caption{
Fragmentation functions used in comparison with the data. For the BCFY function
$f_1(r) = 3(3-4r)$, $f_2(r) = 12-23r+26r^2$, $f_3(r)=(1-r)(9-11r+12r^2)$, and
$f_4(r)=3(1-r)^2(1-r+r^2)$.} 
\end{table}

\begin{table}[t]
\renewcommand{\baselinestretch}{1.0}  
\medskip
\centering
\begin{tabular}{|l|c|c|c|} \hline
Function & $\chi^2$/d.o.f. & parameters & $<x_{E_B}>$ \\
\hline 
Peterson &   14.0/11  &  $\epsilon_b=0.034\pm0.006^*$ & 0.717          \\
\hline
Lund     &   9.6/10   &     $a= 1.7\pm0.2$  & 0.743 \\
         &          &         $b=0.19\pm 0.01$  &    \\
\hline
BCFY      &   22.4/11  &      $r=0.20\pm 0.02$  & 0.705 \\
\hline
NCM      &   15.9/11    &   $\alpha=9\pm2$  & 0.687 \\
         &               &   $\beta=44\pm8$  &  \\ 
\hline
ALEPH    &  9.7/9   & $b=0.0\pm1.0$           & 0.730 \\
         &          & $c=0.78\pm0.05$       &   \\
         &          & $d=0.042\pm0.004$     &       \\
\hline
3$^{rd}$-order polynomial &   14.9/9       &  $b=-7.53\pm0.04$  & -- \\
         &          &  $c=16.49\pm0.07$  &  \\
         &          &  $d=-9.98\pm0.07$   &       \\
\hline
\end{tabular}
\caption{
Results of optimisation of fragmentation functions to the reconstructed B
hadron energy distribution. For the NCM fit the QCD parameters were fixed at
$\Lambda_f$ = 200 MeV and $\mu = m_b = 4.5$ GeV. $^*\,$This value of 
$\epsilon_b$ refers to the B-hadron energy distribution; it should not be 
confused with the value of $\epsilon_b$ used as input in the JETSET
model at the b-quark fragmentation level (Section 5), which 
is significantly lower.}
\end{table}

  
{\footnotesize
\begin{table}[h]
\medskip
\begin{center}
\begin{tabular}{|l|c|c|} \hline
Error source    &     Variation & Error (\%)  \\  \hline 
DETECTOR MODELLING &            &  \\  \hline
Neutral fragmentation energy:   &  &                \\
 \ \ \ \ \ cluster energy scale & $\pm2.2$\% & $^{+0.12}_{-0.27}$ \\ 
 \ \ \ \ \ min. clus energy     & $100^{-0}_{+100}$ MeV & $^{+0.00}_{-0.21} $  
  \\ 
Tracking inefficiency           &   $2.4\mp2.4$\%    & $^{+0.2}_{-1.0}$ \\ 
Lepton mis-ID background        & $\pm25$\%   &  $^{+0.66}_{-0.65}$ \\ \hline
PHYSICS MODELLING               & & \\  \hline
B meson / baryon lifetime  & $1.55\pm0.05$ / $1.10\pm0.08$ ps & 
$^{+0.11}_{-0.12}$ \\ 
B$^{**}$ production             & $20.7 \pm 7$\% & $^{+0.68}_{-0.10}$ \\ 
B$^{**}$ mass                   & $5.704 \pm 0.020$ GeV & $^{+0.03}_{-0.00}$ \\ 
$f^*$ $\equiv$ $\Gamma($B$\rightarrow$D$^*)/\Gamma$(B$\rightarrow$D) & 
${f^*}^{+0}_{-f^*/3}$ & 
$^{+0.32}_{-0.00}$ \\
$f^{**}$ $\equiv$ $\Gamma$(B$\rightarrow$D$^{**})/\Gamma$(B$\rightarrow$D) & 
${f^{**}}\pm f^{**}/3$
 & $^{+0.32}_{-0.21}$ \\ 
B$_u$, B$_d$ / B$_s$ / b-baryon production & $40.1\pm20.0$\% / 
$11.6\pm8.0$\% / $7.0\pm4.0$\% & 
$^{+0.51}_{-0.48}$ \\ 
B$_u$, B$_d$, B$_s$, b-baryon decay modes & $\pm1\sigma$ & $^{+0.11}_{-0.12}$ \\
B-decay charged multiplicity & 5.3$\pm0.2$ tracks & $^{+0.25}_{-0.16}$ \\
$c$-fragmentation: $<x_{E_D}>$ & $0.484\pm0.008$  & $\pm0.01$ \\ 
D$^0$ / D$^+$ / D$_s$ / c-baryon production & 
$56.0\pm5.3\%$ / $23.0\pm3.7\%$ / 
$12.0\pm7.0\%$ / $8.9\pm0.5$\% & $\pm0.01$ \\ 
D decay multiplicity & Ref.~\cite{coffman} & $^{+0.04}_{-0.05}$\\ 
\ss production  & $\pm10$\% & $^{+0.37}_{-0.40}$ \\
$R_b$ & $0.2216\pm0.0010$            & $^{+0.00}_{-0.01}$ \\ 
$R_c$ & $0.16\pm0.01$                & $^{+0.02}_{-0.04}$ \\ 
g $\rightarrow$ \bb splitting & $\pm 50\%$       & $^{+0.23}_{-0.30}$ \\
g $\rightarrow$ \cc splitting & $\pm 50\%$       & $^{+0.22}_{-0.25}$ \\ \hline
Total & & $^{+1.32}_{-1.48}$ \\  \hline
\end{tabular}
\end{center}
\caption{
Systematic errors on $\langle x_{E_B} \rangle$. 
}
\end{table}
}

\begin{table}[t]
\renewcommand{\baselinestretch}{1.0}  
\medskip
\centering
\begin{tabular}{|c|c|c|c|c|} \hline
$x_{E_B}$ bin center  & 1/$\sigma$d$\sigma$/d$x_{E_B}$ & Stat. error & 
Syst. error & Unfolding uncertainty \\
\hline
0.037 & 0.0 &0.0 &0.0 &0.0 \\
0.110 & 0.104 & 0.041 & 0.055 & 0.041 \\
0.183 & 0.105 & 0.050 & 0.068 & 0.035 \\ 
0.256 & 0.158 & 0.076 & 0.095 & 0.043 \\
0.329 & 0.248 & 0.099 & 0.102 & 0.064 \\
0.402 & 0.358 & 0.115 & 0.096 & 0.074 \\
0.475 & 0.560 & 0.136 & 0.095 & 0.061 \\
0.548 & 0.951 & 0.167 & 0.126 & 0.033 \\
0.621 & 1.489 & 0.204 & 0.137 & 0.088 \\
0.694 & 2.136 & 0.242 & 0.164 & 0.171 \\
0.767 & 3.011 & 0.278 & 0.164 & 0.191 \\
0.840 & 2.944 & 0.285 & 0.251 & 0.112 \\
0.913 & 1.460 & 0.211 & 0.319 & 0.144 \\
0.986 & 0.164 & 0.067 & 0.118 & 0.041 \\
\hline
\end{tabular}
\caption{
The fully-corrected scaled B hadron energy distribution.}
\end{table}
  
\clearpage


\section*{Figure Captions}

\noindent
Figure 1:
Candidate D-vertex distributions: (a) number of tracks per vertex; 
(b) vertex mass; 
(c) projection of the vertex flight distance from the IP along the jet 
axis. Candidate B-vertex distributions: (d) vertex mass; (e) projection
along the D-vertex momentum vector
of the vector between the D vertex and the B vertex. 
Data (points with error bars) and simulation (solid histogram); the dashed
histogram shows the simulated contribution from true B \ra D$l$X decays.
In (a) all cuts were applied.
In (b)-(e) all cuts were applied except those on the quantity shown, and 
these latter cut positions (see text) are indicated by arrows.    

\vskip .5truecm

\noindent
Figure 2:
Distribution of (a) electron and (b) muon transverse momentum w.r.t. the jet
axis in jets containing a selected D vertex and respective lepton.
Data (points with error bars) and simulation (histogram). The composition of
the simulated distributions in terms of leptons from B \ra $l$ decays,
cascade B \ra C \ra $l$ decays, wrongly-assigned leptons, promptly
produced C \ra l decays, and fake leptons is indicated.

\vskip .5truecm

\noindent
Figure 3:
The efficiency $\epsilon$ for selecting B hadron decays, as a function of
scaled energy $x_{E_B}$.
Note that the first bin (no point shown) is beneath the kinematic limit 
for $x_{E_B}$.

\vskip .5truecm

\noindent
Figure 4:
Distribution of non-B-associated (a) charged and (b) neutral energy in 
jets containing a candidate B \ra D$l$X decay. 
Data (points with error bars) and simulation (histogram).

\vskip .5truecm

\noindent
Figure 5:
Simulated correlation between the true and reconstructed values of the
non-B-associated (a) neutral and (b) charged energy in 
jets containing a candidate B \ra D$l$X decay. In each bin of 
reconstructed energy the error bar
represents the corresponding r.m.s. deviation in the true energy.
Each line represents a fit to the correlation (see text). 

\vskip .5truecm

\noindent
Figure 6:
Distribution of the normalised difference between the true and reconstructed
B hadron energies in simulated events. The solid line is a fit of the sum of
two Gaussian distributions (see text). The two component Gaussian distributions
are indicated by the dashed lines.

\vskip .5truecm

\noindent
Figure 7:
The distribution of reconstructed scaled energies for B hadron candidates; 
data (points with error bars) and simulation (solid histogram).
Also shown (dashed histogram) is the simulated contribution from non-\bb
events.

\vskip .5truecm

\noindent
Figure 8:
The background-subtracted distribution of reconstructed scaled B hadron 
energy. The data (points with error bars) are compared with simulations
based on six different input B fragmentation functions (see text)
represented by lines joining entries at the bin centers.

\vskip .5truecm

\noindent
Figure 9:
Data distribution of scaled B hadron energy corrected using simulations
based on different input B fragmentation functions (see text): (a) ALEPH, 
(b) Peterson, (c) Lund, (d) BCFY and (e) NCM functions. Statistical error 
bars are shown; these are highly correlated between bins
and among the five sets of results. (f) The five
optimised functional forms used in the correction.

\vskip .5truecm

\noindent
Figure 10:
The final corrected distribution of scaled B hadron energies. In each bin
the statistical error is indicated by the innermost error bar, the 
quadrature sum of
statistical and experimental systematic errors by the middle error bar,  
and the quadrature sum of statistical, experimental systematic and unfolding 
errors by the outermost error bar. Note that the first bin (no point shown) 
is beneath the kinematic limit for $x_{E_B}$.

\vfill\eject

\pagestyle{empty}
 
\begin{figure} [t]
 \hspace*{5cm}
   \epsfxsize=6.0in
    \setlength{\baselineskip}{13pt}
   \begin{center}\mbox{\epsffile{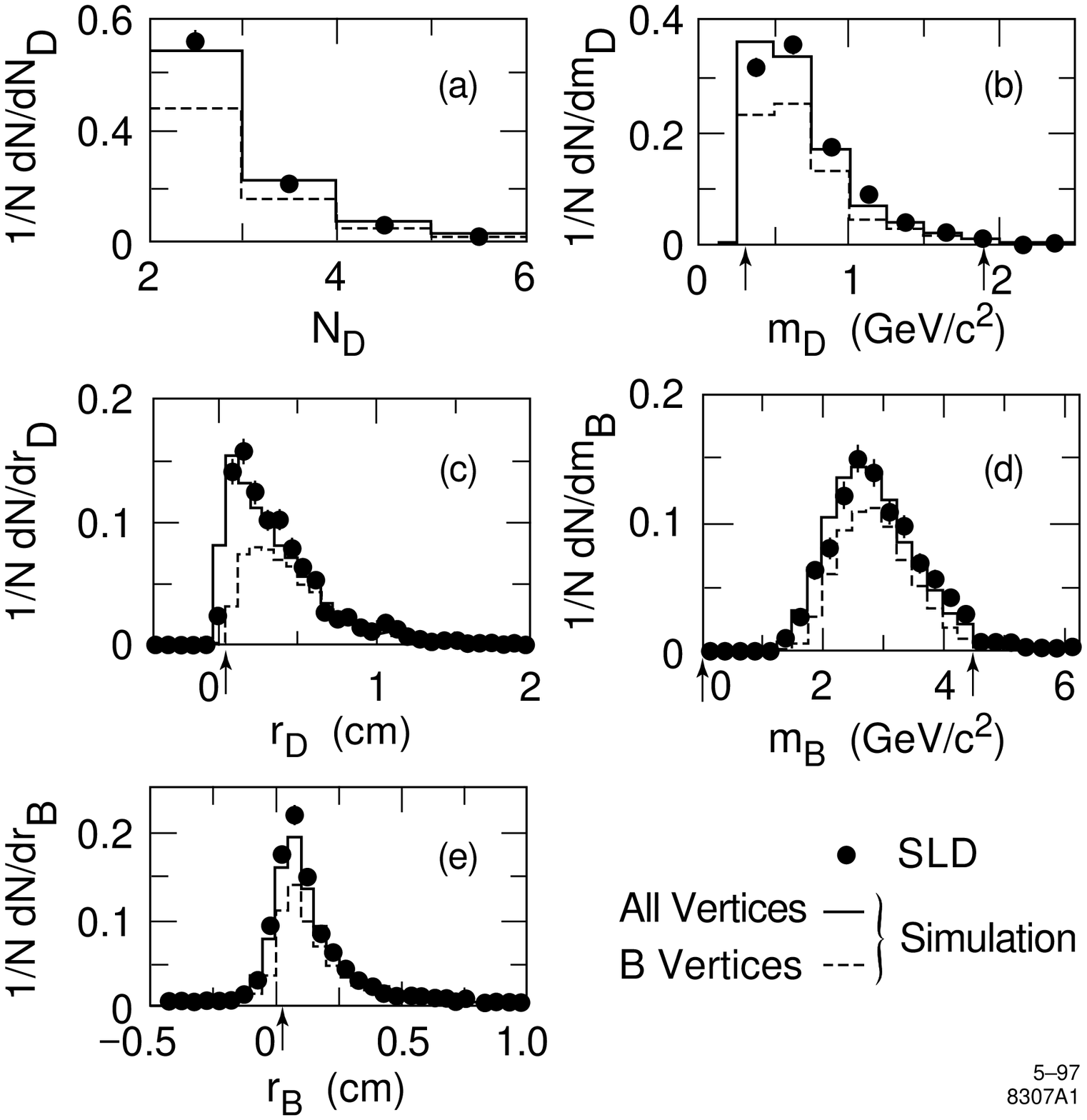}}\end{center}
   \caption[]{}
\end{figure}

\begin{figure} [t]
 \hspace*{5cm}
   \epsfxsize=6.0in
    \setlength{\baselineskip}{13pt}
   \begin{center}\mbox{\epsffile{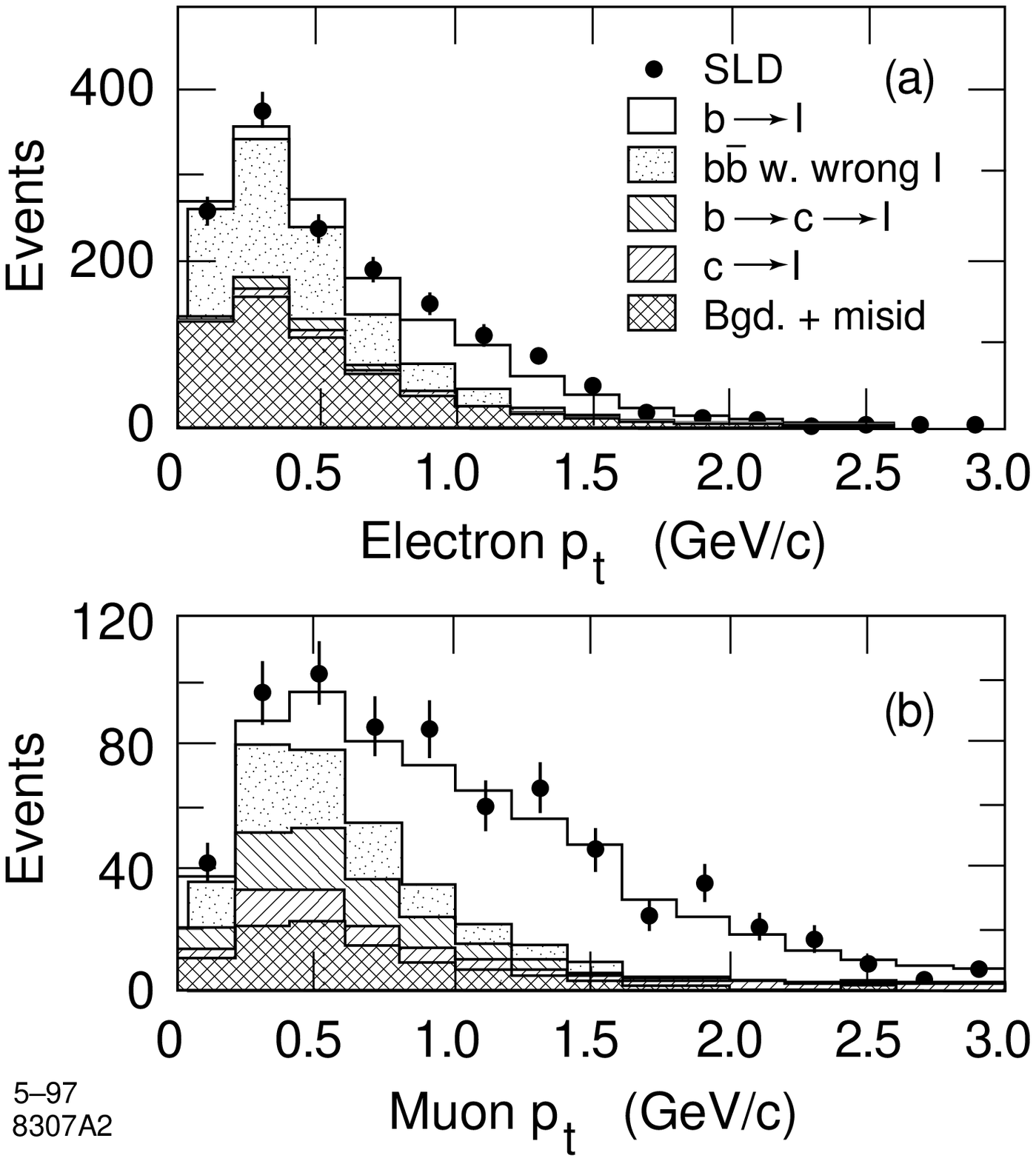}}\end{center}
   \caption[]{}
\end{figure}

\begin{figure} [t]
 \hspace*{5cm}
   \epsfxsize=6.0in
    \setlength{\baselineskip}{13pt}
   \begin{center}\mbox{\epsffile{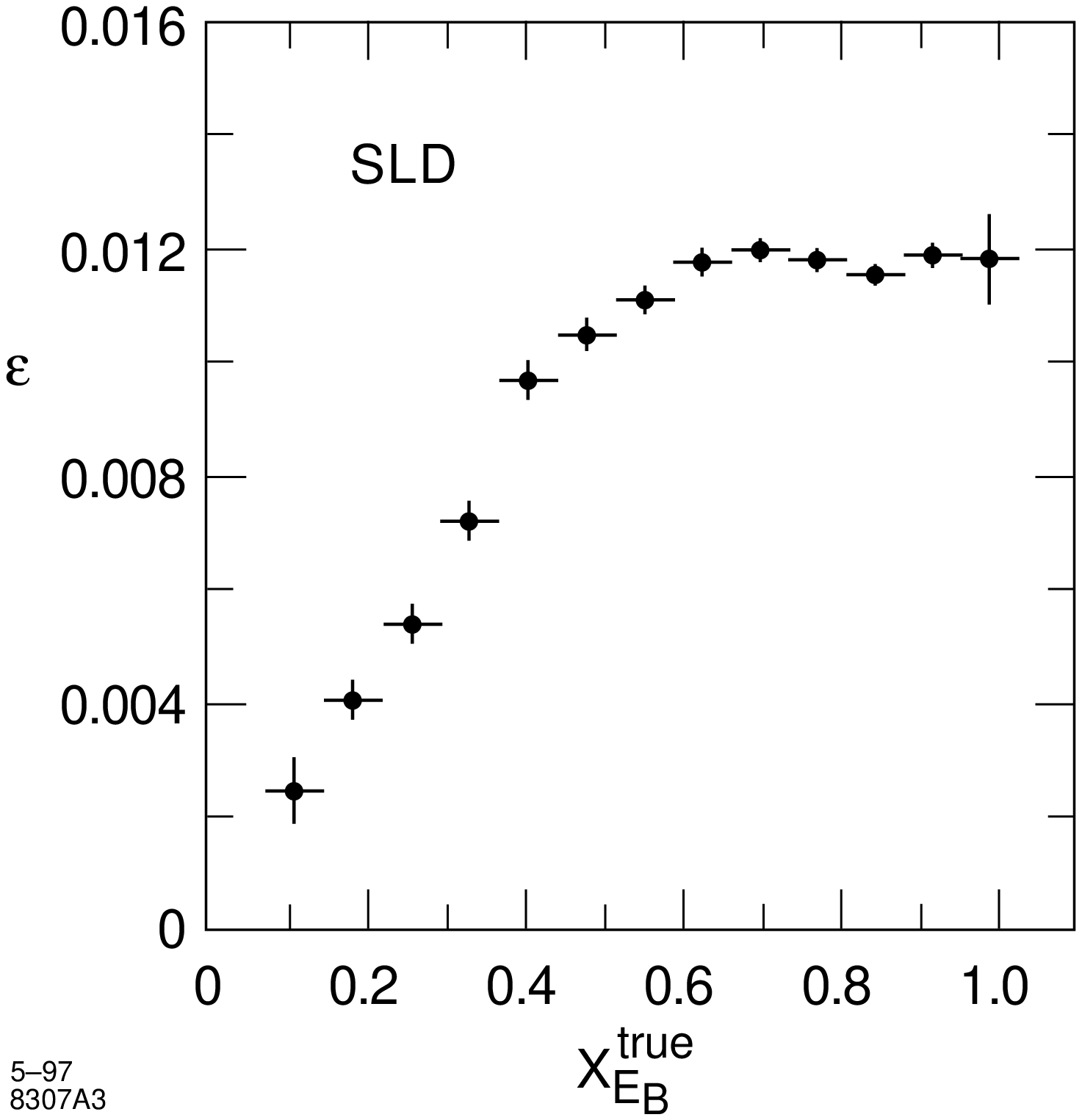}}\end{center}
   \caption[]{}
\end{figure}

\begin{figure} [t]
 \hspace*{5cm}
   \epsfxsize=6.0in
    \setlength{\baselineskip}{13pt}
   \begin{center}\mbox{\epsffile{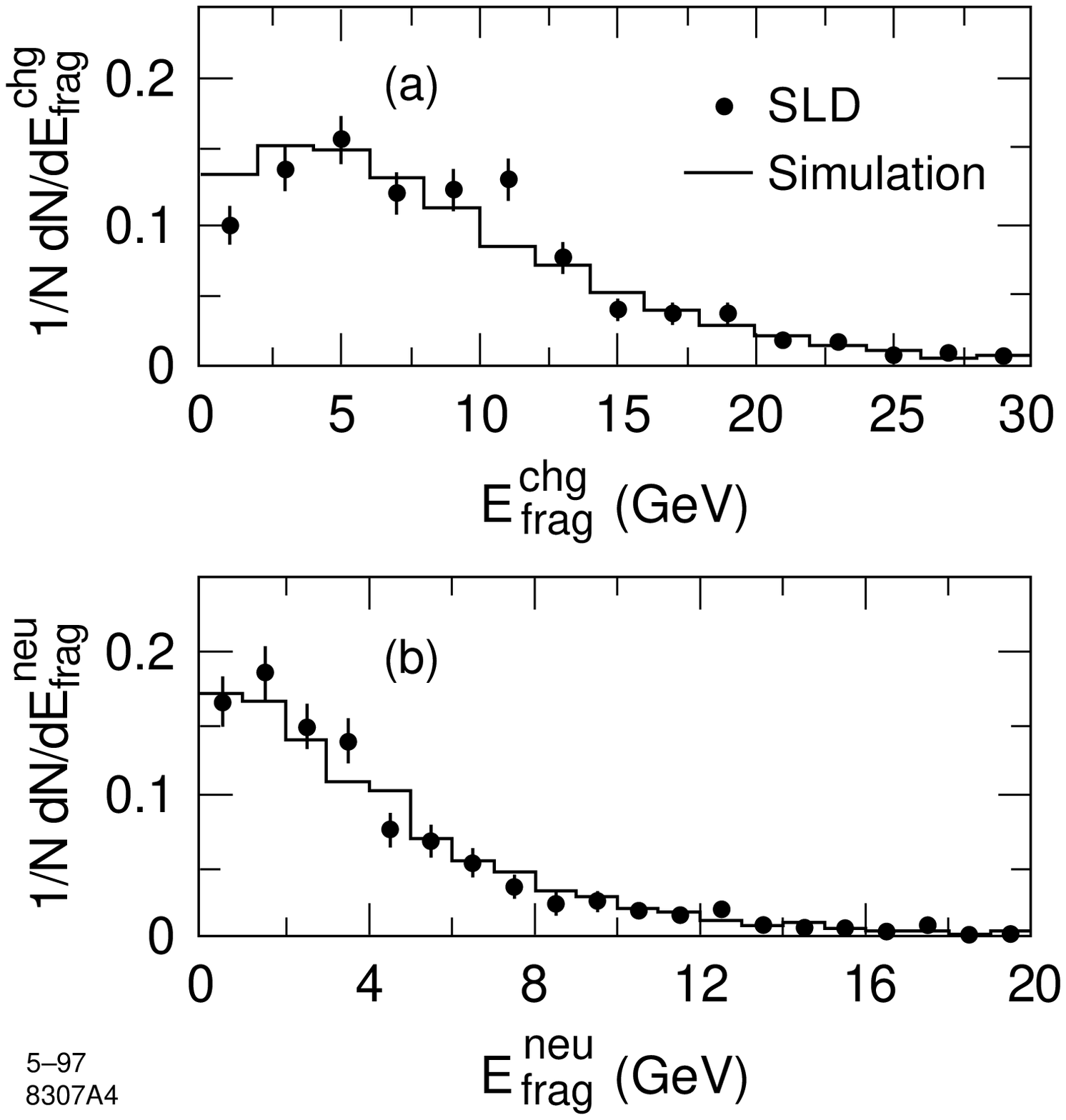}}\end{center}
   \caption[]{}
\end{figure}

\begin{figure} [t]
 \hspace*{5cm}
   \epsfxsize=4.0in
    \setlength{\baselineskip}{13pt}
   \begin{center}\mbox{\epsffile{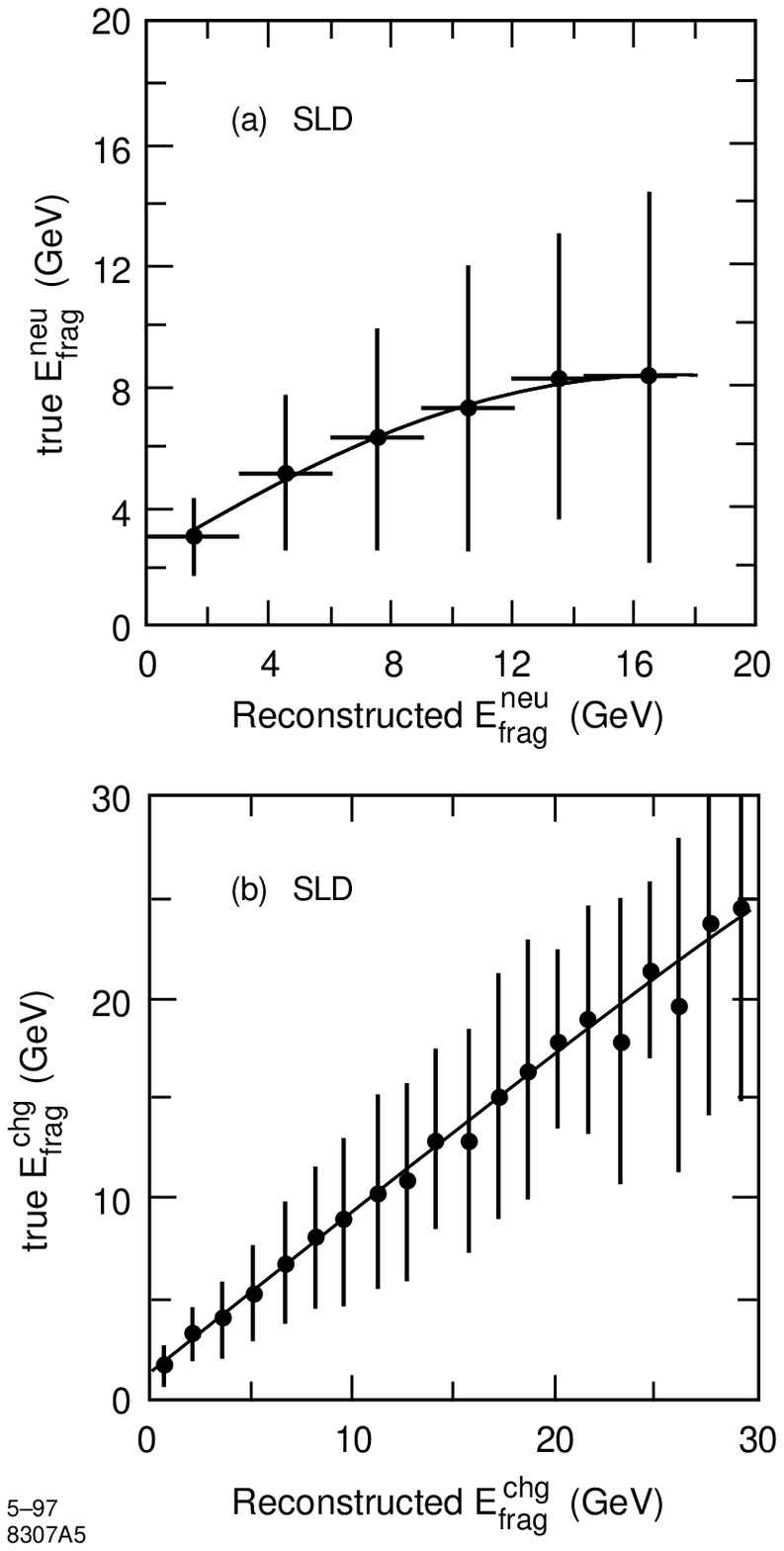}}\end{center}
   \caption[]{}
\end{figure}

\begin{figure} [t]
 \hspace*{5cm}
   \epsfxsize=6.0in
    \setlength{\baselineskip}{13pt}
   \begin{center}\mbox{\epsffile{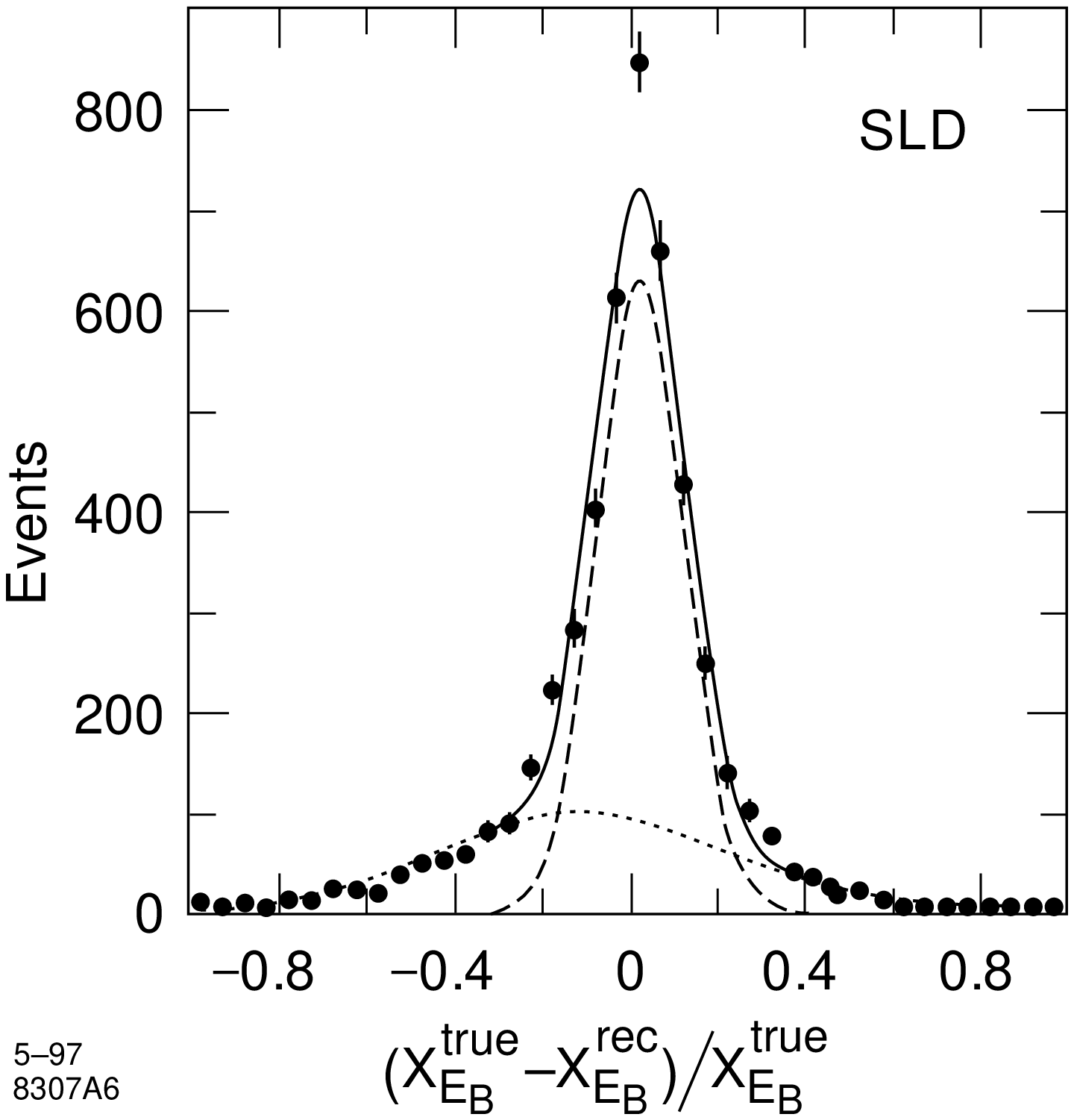}}\end{center}
   \caption[]{}
\end{figure}

\begin{figure} [t]
 \hspace*{5cm}
   \epsfxsize=6.0in
    \setlength{\baselineskip}{13pt}
   \begin{center}\mbox{\epsffile{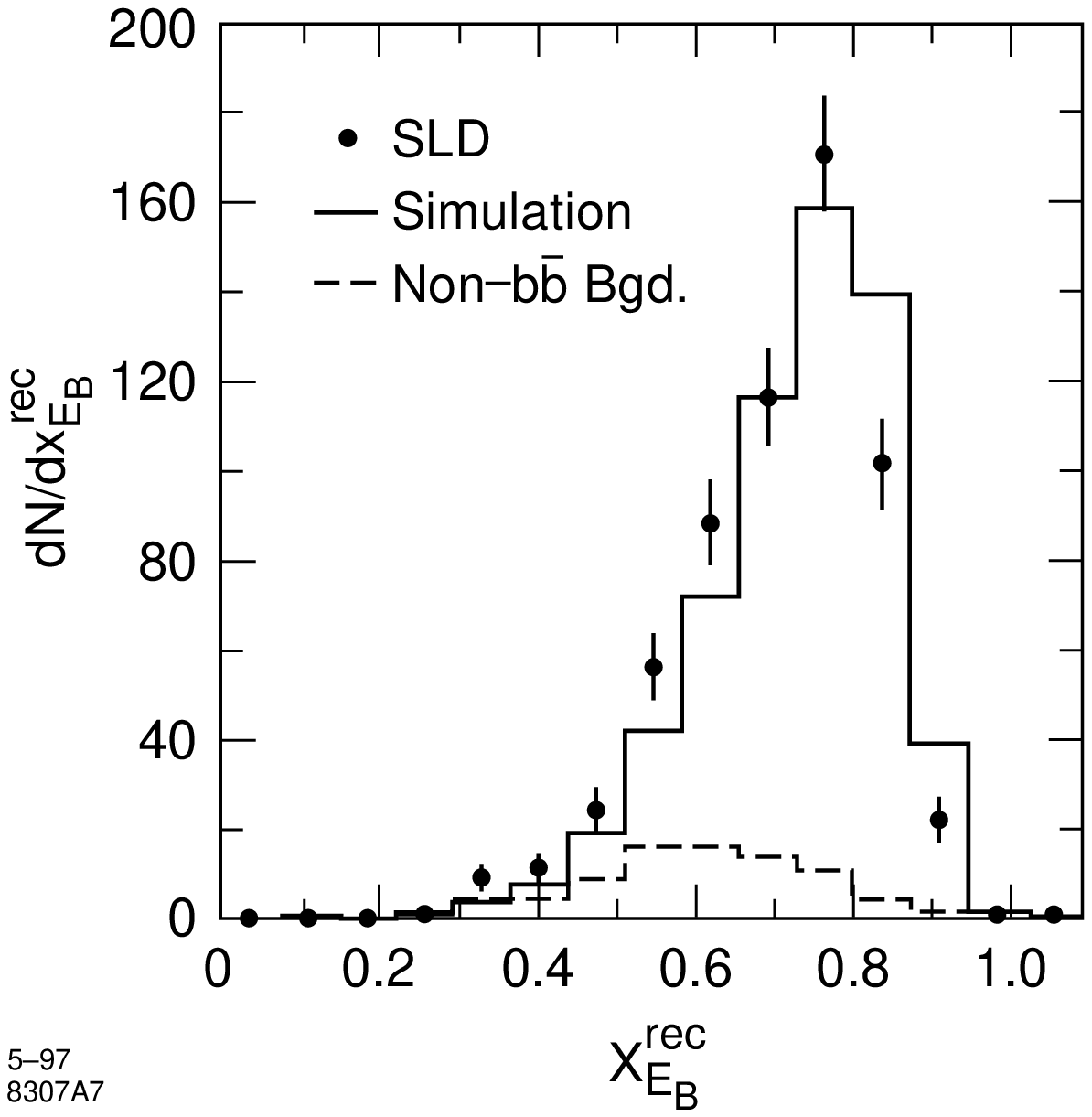}}\end{center}
   \caption[]{}
\end{figure}

\begin{figure} [t]
 \hspace*{5cm}
   \epsfxsize=6.0in
    \setlength{\baselineskip}{13pt}
   \begin{center}\mbox{\epsffile{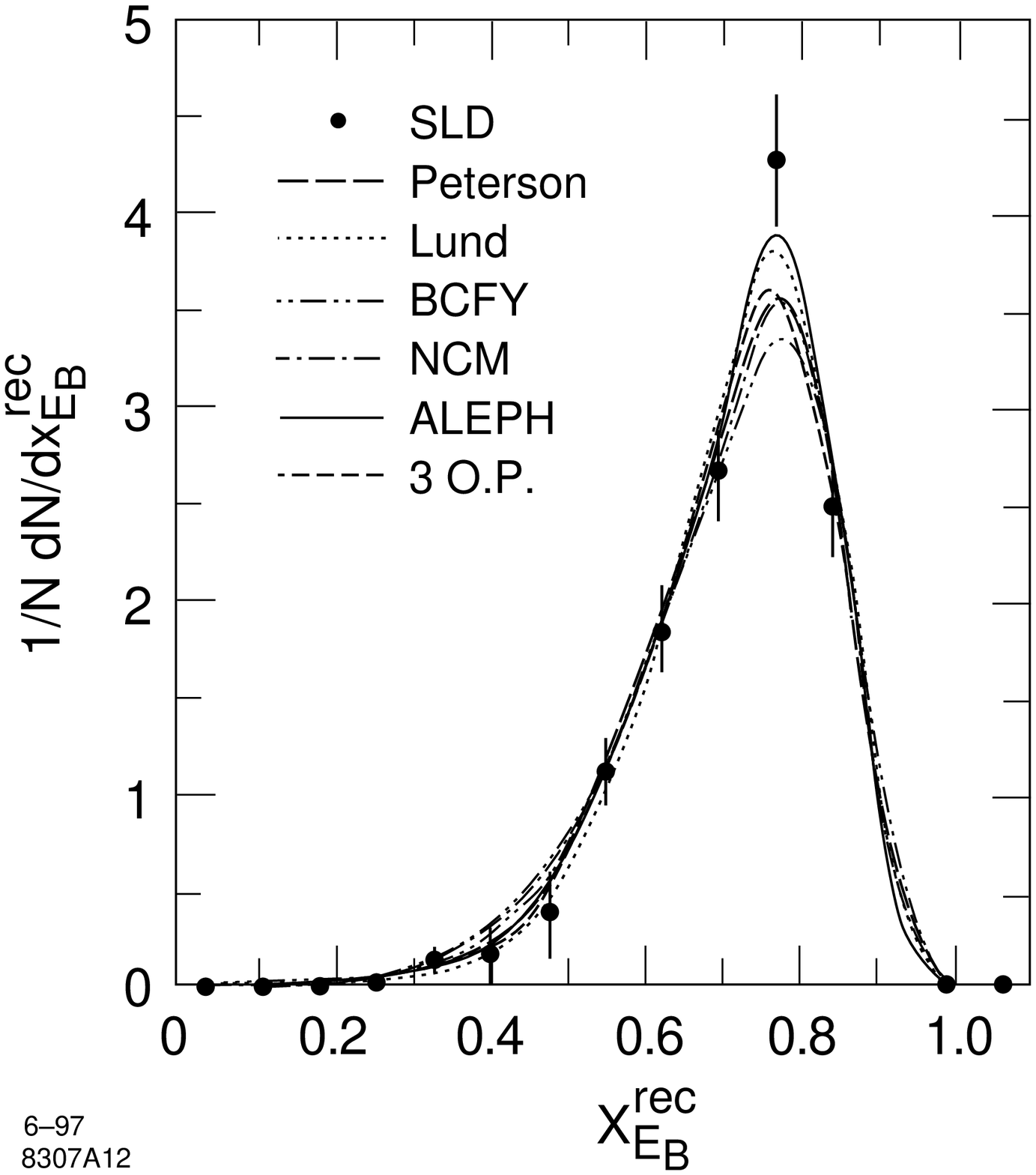}}\end{center}
   \caption[]{}
\end{figure}

\begin{figure} [t]
 \hspace*{5cm}
   \epsfxsize=6.0in
    \setlength{\baselineskip}{13pt}
   \begin{center}\mbox{\epsffile{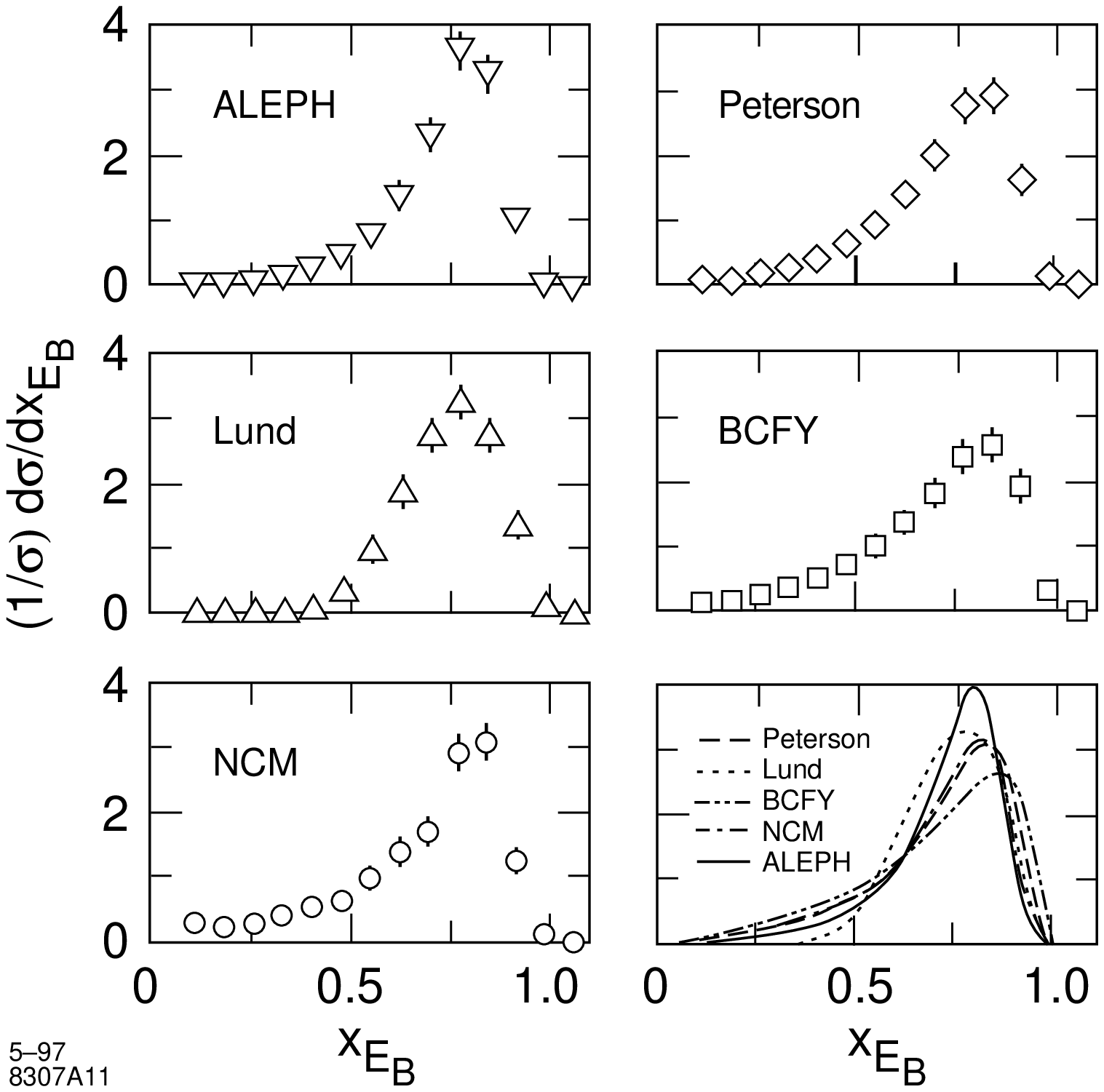}}\end{center}
   \caption[]{}
\end{figure}

\begin{figure} [t]
 \hspace*{5cm}
   \epsfxsize=6.0in
    \setlength{\baselineskip}{13pt}
   \begin{center}\mbox{\epsffile{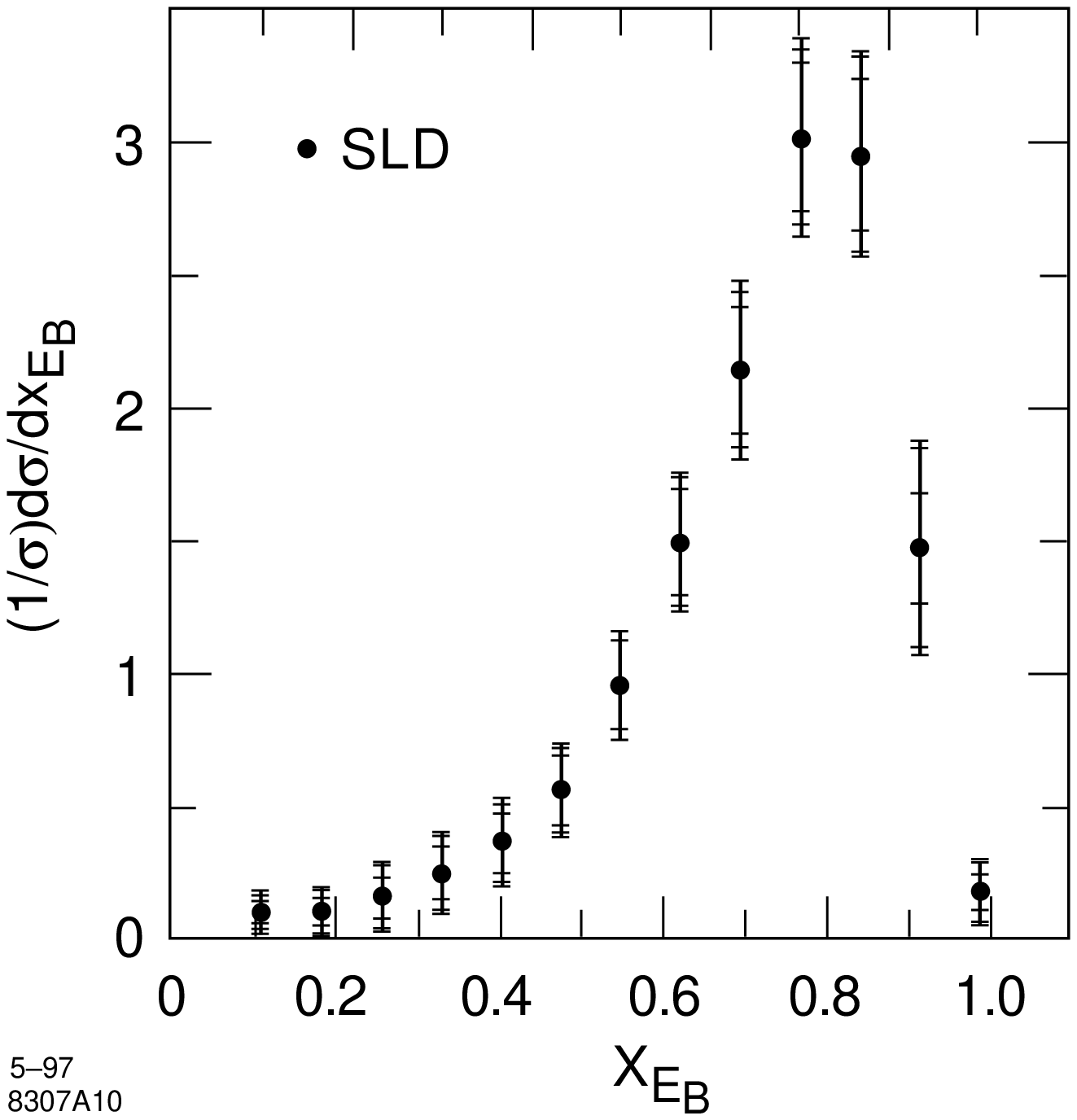}}\end{center}
   \caption[]{}
\end{figure}

\end{document}